\documentclass[a4paper, amsfonts, amssymb, amsmath, reprint, showkeys, nofootinbib, twoside, onecolumn,notitlepage]{revtex4-1}

\def\apj{{ApJ}}                 

\def\mnras{{MNRAS}}             
\def\prb{{Phys.~Rev.~B}}        
\def\prd{{Phys.~Rev.~D}}        
\def\pre{{Phys.~Rev.~E}}        
\def\prl{{Phys.~Rev.~Lett.}}    
\usepackage{amsmath,amstext}
\usepackage[T1]{fontenc}
\usepackage{amssymb}
\input{epsf}
\usepackage{graphicx}
\usepackage{ae,aecompl}

\usepackage{hyperref}
\usepackage{amsmath}
\usepackage{amssymb}
\usepackage{mathtools}
\usepackage{bm}
\usepackage{cleveref}
\usepackage{tensor}
\usepackage{braket}
\usepackage{enumitem}
\usepackage{mhchem}
\usepackage{cancel}
\usepackage{upgreek}

\usepackage{color}

\newcommand{\spc}{\quad \quad \quad}

\def\be{\begin{equation}}
\def\ee{\end{equation}}
\def\beq{\begin{eqnarray}}
\def\eeq{\end{eqnarray}}

\begin{document}
\title{Universality Classes of Relativistic Fluid Dynamics: Applications}
\author{L.~Gavassino$^1$, M.~Disconzi$^{1}$, \& J.~Noronha$^2$}
\affiliation{
$^1$Department of Mathematics, Vanderbilt University, Nashville, TN, USA
\\
$^2$Illinois Center for Advanced Studies of the Universe \& Department of Physics,
University of Illinois at Urbana-Champaign, Urbana, IL 61801-3003, USA
}

\begin{abstract}
Using a formalism that was recently developed in a companion paper, we rigorously prove the equivalence, in the linear regime, of a number of apparently different relativistic hydrodynamic theories proposed in the literature. In particular, we show that Hydro+ is indistinguishable from the Israel-Stewart theory for bulk viscosity, which in turn is indistinguishable from a reacting mixture. The two-fluid model for superfluidity coincides with the Israel-Stewart theory for heat conduction in the limit of infinite conductivity, and this explains why the latter has a second sound. Also, MIS$^*$ is equivalent to the Burgers model for viscoelasticity, and this implies that the former must exhibit an elastic behavior at high frequencies. Additionally, we show that if the degrees of freedom and the conservation laws of a hydrodynamic theory have the same geometric character as those of the Israel-Stewart theory, then such theory must be indistinguishable from the Israel-Stewart theory in the linear regime. This explains why all second-order theories turn out to be identical near equilibrium. Finally, we construct the first linearized model for a relativistic supersolid that is proven to be causal, stable, and strongly hyperbolic.
\end{abstract}

\maketitle

\section{Introduction}

In what ways is atmospheric air different from liquid water? What are the differences between liquid water and an electron-positron plasma?
How are electron-positron plasmas different from superfluid helium? Depending on whom you ask (and their background), the answers may be wildly different. For example, one may say that these systems have different chemical compositions, different characteristic temperatures and pressures, different optical properties, the presence or not of a $U(1)$ broken symmetry, and much more. However, let us examine these questions from a purely hydrodynamic perspective. There is certainly a regime where atmospheric air and liquid water are in practice very similar, and in this case one would describe both of them using the same theory, i.e, the Navier-Stokes hydrodynamic equations \cite{landau10}. On the other hand, an electron-positron plasma is fundamentally different from the latter as one needs two velocities to describe it, not just one. This forces us to model the two-component plasma using an intrinsically different hydrodynamic theory (e.g. multifluid hydrodynamics \cite{prix2004}). However, consider now the fact that the standard macroscopic description of a superfluid involves two velocities \cite{landau9}. Does this mean that the hydrodynamic regime of a superfluid can be ``similar'' to that of a two-component plasma just like water can be ``similar'' to air? Indeed, the macroscopic degrees of freedom are essentially the same, but a superfluid differs from a two-component plasma by its conservation laws. As a matter of fact, a superfluid has an additional vector-type conservation law, a triplet of topological winding numbers \cite{AndreevMelnikovsky2004,Termo}, which must be accounted for in a hydrodynamic model, being responsible for the long lifetime of superfluid currents. From a hydrodynamic point of view, this conservation law is ultimately the only ingredient that distinguishes superfluid helium from an electron-positron plasma.

In a companion paper \cite{paperI}, henceforth called Paper I, we showed how the qualitative observations made above can be made quantitative. We determined what are the mechanical properties of hydrodynamic models near equilibrium that are compatible with relativity, for a given collection of degrees of freedom and a given set of conservation laws. This was done through the combination of our Theorems 1, 2, and 3 in \cite{paperI}, which showed that the second law of thermodynamics and thermodynamic stability criteria are such powerful constraints in relativity that the very structure of the linearized fluid equations around equilibrium is fully determined by the choice of degrees of freedom and conservation laws. This implies that all thermodynamically consistent theories of relativistic fluids must fall into new types of universality classes near equilibrium. Each class describes a physically distinct behavior, which is determined by the degrees of freedom, the corresponding information current \cite{GavassinoGibbs2021,GavassinoCausality2021}, and the conservation laws. The existence of such universality classes makes manifest a number of surprising equivalences between systems described by seemingly different sets of equations of motion. 

In this paper we fully explore the physical content of such universality classes and the equivalence between a number of hydrodynamic models. Our analysis is restricted to linear deviations about homogeneous and isotropic equilibria.
Theorem 1 of Ref.\ \cite{paperI} states that if the equilibrium state is thermodynamically stable, i.e. if (calling the linear perturbation fields ``$\varphi^A$'')
\begin{itemize}
    \item [(i)] It is the maximum entropy state for fixed values of the integrals of motion, which means that there is a quadratic information current $E^\mu(\varphi^A)$ that is timelike future directed for non-vanishing linear perturbations \cite{GavassinoGibbs2021,GavassinoCausality2021}, and
    \item[(ii)] The second law of thermodynamics is valid, which means that $\partial_\mu E^\mu=\sigma$, where $\sigma(\varphi^A)\leq 0$ is also a quadratic function of the perturbations,
\end{itemize}
then, under very general mathematical assumptions, the linearized equations of the theory are symmetric hyperbolic \cite{Kato2009} and causal \cite{Wald}, and they take the form
\begin{equation}\label{Fudamental}
\partial_\mu \dfrac{\partial (TE^\mu)}{\partial \varphi^A} = -\dfrac{1}{2} \dfrac{\partial(T \sigma)}{\partial \varphi^A} \, ,
\end{equation}
where the factor $T=\text{constant}$ (the background temperature) is usually included for convenience.

Working in the equilibrium global rest frame, in Paper I \cite{paperI}
we have introduced a classification of relativistic hydrodynamic theories in terms of the transformation properties of their degrees of freedom and conserved densities under the  $SO(3)$ group describing the spatial isotropy of the equilibrium state. In particular, we say that a theory is ``of class $(\mathfrak{g}_0,\mathfrak{g}_1,\mathfrak{g}_2)-(\bar{\mathfrak{g}}_0,\bar{\mathfrak{g}}_1,\bar{\mathfrak{g}}_2)$'' if its dynamical fields $\varphi^A$ can be grouped into $\mathfrak{g}_0$ rotation scalars (such as the temperature, the chemical potential, and the bulk stress), $\mathfrak{g}_1$ rotational vectors (such as the flow velocity, and the heat flux), and $\mathfrak{g}_2$ symmetric traceless tensors with two indices (such as the shear stress tensor), and analogously the zeroth components of the conserved currents $j^\mu_I$ can be grouped into $\bar{\mathfrak{g}}_0$ rotational scalars (such as the energy density, and the baryon density), $\bar{\mathfrak{g}}_1$ rotational vectors (such as the momentum density and the superfluid winding density \cite{AndreevMelnikovsky2004,Termo}) and $\bar{\mathfrak{g}}_2$ symmetric traceless tensors with two indices (such as the elastic deformation, see our companion paper \cite{paperI}). This classification can be naturally extended to include other tensor categories. This paper is organized as follows:

\begin{itemize}
\item In Section \ref{sec:hydroplus}, we use the theorems of Ref.\ \cite{paperI} to show that the theory called ``Hydro+'', developed in \cite{Hydro+2018} as a model for relativistic fluids near the critical point, is symmetric hyperbolic in the linear regime, and it is equivalent to the Israel-Stewart theory for bulk viscosity.
\item In Section \ref{sec:abacadabra}, we provide three examples of apparently different fluid models which turn out to be equivalent in the linear regime, describing types of matter with the same ``mechanical properties''. In particular, we show that a fluid mixture undergoing a chemical reaction is equivalent to a bulk viscous fluid, a superfluid is equivalent to a heat-conducting Israel-Stewart fluid in the limit of infinite heat conductivity, and the Burgers model for viscoelasticity is equivalent to the MIS$^*$ theory proposed in  \cite{KeYin2022}.
\item In Section \ref{sec:apriori}, we show that the linearized Israel-Stewart theory at finite chemical potential \cite{Israel_Stewart_1979} is the most general (thermodynamically consistent) theory of class $(3,2,1)-(2,1,0)$. This finally explains why all second-order theories \cite{Hishcock1983,OlsonLifsh1990,GerochLindblom1990,
carter1991,Baier2008,Denicol2012Boltzmann,
Stricker2019} turn out to be equivalent to the Israel-Stewart theory in the linear regime.
\item In Section \ref{sec:supersolid}, we use the information current to explicitly construct a relativistic linear theory for supersolid hydrodynamics that is causal, stable, and strong hyperbolic, as a representative of the universality class $(3,2,1)-(3,2,1)$.
\item Section \ref{conclusions} presents our conclusions and outlook while Appendix \ref{AAA} describes in more detail the bulk viscous behavior of Hydro+. 
\end{itemize}
We remark that, when in this article we say that two theories are ``equivalent'', we actually mean ``mathematically equivalent'', i.e. there is a change of variables that maps the equations of one theory into the equations of the other theory. Clearly, it may be the case that the equations were originally thought to describe different physical systems (e.g., a superfluid and a heat-conductive fluid). Furthermore, it may also happen that the two theories were constructed having different domains of applicability in mind, or different assumptions about the origin and interpretation of the transport coefficients. For example, in the following, we will prove that Hydro+ is mathematically equivalent to the Israel-Stewart theory for bulk viscosity. However, Hydro+ \cite{Hydro+2018} was constructed having in mind only systems with a \textit{parametrically slow} relaxing mode, while Israel-Stewart (at least in its ``canonical'' derivations  \cite{Israel_Stewart_1979,Hishcock1983,Baier2008,DMNR2012}) makes no reference to a timescale separation of this kind. Our method is blind to these physical details. In a sense, this is one of the main messages of the paper: physically different systems may happen to be described by identical equations and, thus, exhibit the same mechanical properties. Also, we would like to stress again that our equivalences hold only at the level of linear theory, i.e., for the equations of motion linearized about homogeneous equilibrium states.

We use natural units $\hbar=k_B = c=1$ in Minkowski spacetime with Cartesian coordinates and metric signature $(-,+,+,+)$. Greek indices run from 0 to 3, while lowercase Latin indices run from 1 to 3. Uppercase Latin indices are multi-index labels.

\section{Hydro+ belongs to the bulk viscosity class}\label{sec:hydroplus}

Our first example serves mostly an illustrative purpose: we will use it as a test case to show how the formalism can be used to systematically analyze the mathematical structure of a theory in the linear regime. Our ``test theory'' will be the Hydro+ theory developed in  \cite{Hydro+2018}. We will follow the steps below:
\begin{itemize}
\item First we will compute the information current explicitly. 
\item Then, we will use Theorem 1 of our companion paper to prove that, in the linear regime, Hydro+ is symmetric hyperbolic and causal, provided that it is thermodynamically stable.
\item Finally, we will show that the equations of Hydro+ are mathematically equivalent to those of the Israel-Stewart theory for bulk viscosity, in the linear regime. Indeed, it cannot be otherwise, since they both belong to the same universality class, namely $(3,1,0)-(2,1,0)$.
\end{itemize}

\subsection{A quick overview of Hydro+}

The algebraic degrees of freedom of Hydro+ are $\psi^A=\{\rho,n,u^\mu,\phi \}$, namely energy density, baryon density, flow velocity ($u_\mu u^\mu=-1$), and an additional non-conserved degree of freedom \cite{Hydro+2018}. The theory postulates the existence of a non-equilibrium entropy density $s=s(\rho,n,\phi)$, whose differential reads
\begin{equation}\label{dsHydroplus}
ds=\beta \, d\rho -\alpha \, dn - \lambda \, d\phi \, ,
\end{equation}
where $\beta$, $\alpha$, and $\lambda$ are the inverse temperature, the fugacity, and the ``$\phi$-affinity'' \cite{PrigoginebookModernThermodynamics2014}. A distinctive feature of Hydro+ is that the formula for the pressure, $P=P(\rho,n,\phi)$, is not uniquely determined in terms of $s(\rho,n,\phi)$ but it must be provided independently. Fixing the functions $s(\rho,n,\phi)$ and $P(\rho,n,\phi)$, the constitutive relations take the usual perfect fluid form (we neglect gradient corrections, which would increase the number of dynamical degrees of freedom \cite{GavassinoLyapunov_2020}): 
\begin{equation}
\begin{split}
T^{\mu \nu}={}& (\rho+P)u^\mu u^\nu+Pg^{\mu \nu} \, ,\\
n^\mu ={}& nu^\mu \, ,\\
s^\mu ={}& su^\mu \, ,\\
\end{split}
\end{equation}
where $T^{\mu \nu}$, $n^\mu$, and $s^\mu$ are the energy-momentum tensor, the baryon current, and the entropy current. The equations of motion are the conservation laws, $\partial_\mu T^{\mu \nu}=0$, $\partial_\mu n^\mu=0$, and the entropy production equation $\partial_\mu s^\mu=\lambda F_\phi \geq 0$, where $F_\phi(\rho,n,\phi)$ is a transport coefficient, called the ``returning force''. Combining these equations, we find
\begin{equation}\label{hydrolpusfieldmode}
u^\mu \partial_\mu \phi =-A_\phi \partial_\mu u^\mu -F_\phi \, ,
\end{equation}
where $A_\phi(\rho,n,\phi)$ is a ``compressibility'' defined by the relation $\lambda A_\phi =\beta (\rho+P)-s-\alpha n$. Note that, if one is given the functions $s(\rho,n,\phi)$ and $A_\phi(\rho,n,\phi)$, they can be used to reconstruct the equation of state for the pressure, $P=P(\rho,n,\phi)$. Also, when $\lambda=0$, $s$ and $P$ must be related by the usual Euler relation $s=\beta(\rho+P)-\alpha n$.

\subsection{Computation of the information current}

To compute the information current, we follow the standard procedure \cite{GavassinoCausality2021,
GavassinoStabilityCarter2022,GavassinoGENERIC2022}. We imagine to bring the fluid into weak contact with a non-rotating thermal reservoir which moves with four-velocity $u_\nu^\star$, has temperature $T^\star$ and baryon chemical potential $\mu^\star$. These parameters are treated as constants,  because the reservoir is infinitely larger than the fluid. We look for the equilibrium state by maximizing (at constant $u_\nu^\star$, $T^\star$, $\mu^\star$) the function $\Phi {=}\int \Phi^\mu d\Sigma_\mu$, with
\begin{equation}
\Phi^\mu = s^\mu +\dfrac{\mu^\star}{T^\star} n^\mu +\dfrac{u^\star_\nu}{T^\star} T^{\mu \nu} \, .
\end{equation}
In order to do this, we consider an arbitrary smooth one-parameter family of states $\psi^A(\epsilon)$, where $\epsilon=0$ is the equilibrium state, so that $\dot{\Phi}^\mu(0)=0$ (notation: $\dot{f}=df/d\epsilon$). This gives us the equilibrium conditions for the theory. Then, the perturbation fields are defined as $\varphi^A=\dot{\psi}^A(0)$, and the information current is just $E^\mu =-\ddot{\Phi}^\mu(0)/2$. Explicitly, one finds:
\begin{equation}
\begin{split}
T^\star \Phi^\mu ={}& \big[ T^\star s+\mu^\star n + u^\star_\nu u^\nu (\rho+P) \big]u^\mu +P u^{\star \mu} \, ,\\
T^\star \dot{\Phi}^\mu ={}& \big[ T^\star \dot{s}+\mu^\star \dot{n} + u^\star_\nu \dot{u}^\nu (\rho+P) +u^\star_\nu u^\nu (\dot{\rho}+\dot{P}) \big]u^\mu+ \big[ T^\star s+\mu^\star n + u^\star_\nu u^\nu (\rho+P) \big]\dot{u}^\mu +\dot{P} u^{\star \mu} \, ,\\
T^\star \ddot{\Phi}^\mu ={}& \big[ T^\star \ddot{s}+\mu^\star \ddot{n} + u^\star_\nu \ddot{u}^\nu (\rho+P) +u^\star_\nu u^\nu (\ddot{\rho}+\ddot{P})+2u^\star_\nu \dot{u}^\nu (\dot{\rho}+\dot{P}) \big]u^\mu+ \big[ T^\star s+\mu^\star n + u^\star_\nu u^\nu (\rho+P) \big]\ddot{u}^\mu \\
& +2\big[ T^\star \dot{s}+\mu^\star \dot{n} + u^\star_\nu \dot{u}^\nu (\rho+P) +u^\star_\nu u^\nu (\dot{\rho}+\dot{P}) \big]\dot{u}^\mu +\ddot{P} u^{\star \mu} \, ,\\
\end{split}
\end{equation}
The stationarity requirement $\dot{\Phi}^\mu(0)=0$ produces the usual conditions for thermodynamic equilibrium: $\beta^{-1}=T^\star$, i.e. the fluid and reservoir have the same temperature; $T^\star\alpha=\mu^\star$, i.e. the fluid and reservoir have the same chemical potential; and $u_\nu = u^\star_\nu$, i.e. the fluid is comoving with the reservoir (in agreement with the zeroth law of thermodynamics \cite{GavassinoTermometri}). Additionally, we have that $\lambda=0$, in agreement with \cite{Hydro+2018}. Plugging these conditions in the equation for $\ddot{\Phi}^\mu(0)$, and invoking the identities
\begin{equation}
\begin{split}
u_\nu \ddot{u}^\nu={}& -\dot{u}_\nu \dot{u}^\nu \, , \\
\ddot{s}^\mu ={}& \beta \ddot{\rho}-\alpha \ddot{n}-\lambda \ddot{\phi}+\dot{\beta}\dot{\rho}-\dot{\alpha}\dot{n}-\dot{\lambda}\dot{\phi} \, ,\\
\end{split}
\end{equation} 
where the first follows from the normalization $u_\nu u^\nu=-1$ and the second follows from \eqref{dsHydroplus}, we finally obtain
\begin{equation}
E^\mu = \dfrac{1}{2} \big[ -\delta \beta \,\delta \rho+\delta \alpha \, \delta n+\delta \lambda \,\delta \phi +\beta(\rho+P) \delta u_\nu \, \delta u^\nu \big]u^\mu+ \beta \delta P \,\delta u^\mu \, ,
\end{equation}
where we have introduced the notation $\delta f := \dot{f}(0)$. For Hydro+ to be stable  
in all reference frames \cite{Hiscock_Insatibility_first_order,
GavassinoFronntiers2021,
GavassinoSuperluminal2021,GavassinoBounds2023}, $E^\mu$ should be future directed non-spacelike \cite{GavassinoCausality2021}.

\subsection{Linearized Hydro+ is symmetric hyperbolic}

It is convenient to treat $\tilde{\varphi}^B=\{\delta \beta, \delta \alpha, \delta \lambda,\delta u^\nu\}$ as the fundamental degrees of freedom of the linear perturbations. Hence, let us introduce the Massieu function $\mathcal{M}:=s-\beta \rho + \alpha n+\lambda \phi$ \cite{Callen_book}, whose differential reads
\begin{equation}
d\mathcal{M}=-\rho \, d\beta +n \, d\alpha +\phi \, d\lambda \, .
\end{equation}
Its Hessian matrix $\mathcal{M}_{BC}$ allows us to express the ``old'' variables ($\delta \rho$, $\delta n$, and $\delta \phi$) in terms of the ``new'' ones ($\delta \beta$, $\delta \alpha$, and $\delta \lambda$). For example, we can write $\delta \phi=\delta (\partial \mathcal{M}/\partial \lambda)=\mathcal{M}_{\lambda\beta} \delta \beta +\mathcal{M}_{\lambda\alpha} \delta \alpha+\mathcal{M}_{\lambda\lambda} \delta \lambda$. Furthermore, if we differentiate the equation $\lambda A_\phi =\beta (\rho+P)-s-\alpha n$ with respect to $\epsilon$, and we evaluate the result at $\epsilon=0$, we obtain that 
\begin{equation}\label{deppummo}
\beta \delta P =-(\rho+P) \,\delta \beta+n \,\delta \alpha+A_\phi \, \delta \lambda \, .
\end{equation}
Finally, since $\partial_\mu s^\mu =\lambda F_\phi$ must be non-negative, we can set $F_\phi=\Xi\lambda$, with $\Xi \geq 0$, and the entropy production rate takes the form $\partial_\mu s^\mu = \Xi \lambda^2$. Combining everything together, and working in the equilibrium rest frame, we obtain
\begin{equation}
\begin{split}
E^0 ={}& \dfrac{1}{2} 
\begin{pmatrix}
\delta \beta, \delta \alpha, \delta \lambda
\end{pmatrix} 
\begin{bmatrix}
\mathcal{M}_{\beta \beta} & \mathcal{M}_{\beta \alpha} & \mathcal{M}_{\beta \lambda} \\ 
\mathcal{M}_{\alpha \beta} & \mathcal{M}_{\alpha \alpha} & \mathcal{M}_{\alpha \lambda} \\ 
\mathcal{M}_{\lambda \beta} & \mathcal{M}_{\lambda \alpha} & \mathcal{M}_{\lambda \lambda} \\ 
\end{bmatrix}
\begin{pmatrix}
\delta \beta \\
\delta \alpha \\
\delta \lambda \\
\end{pmatrix}
+\dfrac{1}{2}\beta(\rho+P) \delta u^j \delta u_j , \\
E^j ={}& -(\rho+P) \,\delta \beta \delta u^j +n \,\delta \alpha \delta u^j +A_\phi \, \delta \lambda \delta u^j \, , \\
\sigma ={}& \Xi (\delta \lambda)^2 \, . \\
\end{split}
\end{equation}
This allows us to write the system \eqref{Fudamental} explicitly for Hydro+, and the result is the following:
\begin{equation}\label{carismatico}
\begin{split}
& \partial_t [\mathcal{M}_{\beta\beta} \delta \beta +\mathcal{M}_{\beta\alpha} \delta \alpha+\mathcal{M}_{\beta\lambda} \delta \lambda]-(\rho+P) \partial_j \delta u^j = 0 \, , \\
& \partial_t [\mathcal{M}_{\alpha\beta} \delta \beta +\mathcal{M}_{\alpha\alpha} \delta \alpha+\mathcal{M}_{\alpha\lambda} \delta \lambda]+n \partial_j \delta u^j = 0 \, , \\
& \partial_t [\mathcal{M}_{\lambda\beta} \delta \beta +\mathcal{M}_{\lambda\alpha} \delta \alpha+\mathcal{M}_{\lambda\lambda} \delta \lambda]+A_\phi \partial_j \delta u^j = -\Xi \delta \lambda \, , \\
& \beta (\rho+P) \partial_t \delta u_k +\partial_k [-(\rho+P) \,\delta \beta+n \,\delta \alpha+A_\phi \, \delta \lambda]=0 \, . \\
\end{split}
\end{equation}
In accordance with Theorem 1 of \cite{paperI}, we see that the above system correctly describes the linearized dynamics of Hydro+. In particular, the first equation is the conservation of energy, $\partial_t \delta \rho+(\rho+P)\partial_j \delta u^j =0$, the second is the conservation of the baryon current, $\partial_t \delta n +n \partial_j \delta u^j=0$, the third is the linearized version of Eq.\ \eqref{hydrolpusfieldmode}, $\partial_t \delta \phi+A_\phi \partial_j \delta u^j = -\Xi \delta \lambda$, and the fourth is the conservation of momentum, $\beta(\rho+P)\partial_t \delta u_k +\beta\partial_k  \delta P =0$. Since the Hessian of $\mathcal{M}$ is symmetric, system \eqref{carismatico} is manifestly symmetric. If the fluid is thermodynamically stable (i.e. $E^\mu$ is future directed non-spacelike), system \eqref{carismatico} is automatically strongly hyperbolic and causal. 

\subsection{Mapping Hydro+ into Israel-Stewart theory}\label{hydrIS}

Hydro+ belongs to the universality class $(3,1,0)-(2,1,0)$, which is the same universality class of the Israel-Stewart theory for bulk viscosity. Hence, there must exist a linear transformation of variables that converts the Hydro+ equations into to the Israel-Stewart equations. Let's verify this explicitly. The linearized Hydro+ field equations are
\begin{equation}\label{comebemybaby}
\begin{split}
& \partial_t \delta \rho = -(\rho+P)\partial_j \delta u^j \, , \\
& \partial_t \delta n = -n\partial_j \delta u^j \, , \\
&(\rho+P)\partial_t \delta u_k +\partial_k  \delta P =0 \, \\
& \partial_t \delta \phi+A_\phi \partial_j \delta u^j = -\Xi \delta \lambda \, . \\
\end{split}
\end{equation}
The first two equations already coincide with the Israel-Stewart ones. Let us decompose $\delta P$ and $\delta \phi$ as linear combinations of $\delta \rho$, $\delta n$, and $\delta \lambda$:
\begin{equation}
\begin{split}
& \delta P =\dfrac{\partial P}{\partial \rho}\bigg|_{n,\lambda} \delta \rho + \dfrac{\partial P}{\partial n}\bigg|_{\rho,\lambda} \delta n +\dfrac{\partial P}{\partial \lambda}\bigg|_{n,\rho} \delta \lambda \, , \\
& \delta \phi =\dfrac{\partial \phi}{\partial \rho}\bigg|_{n,\lambda} \delta \rho + \dfrac{\partial \phi}{\partial n}\bigg|_{\rho,\lambda} \delta n +\dfrac{\partial \phi}{\partial \lambda}\bigg|_{n,\rho} \delta \lambda \, . \\
\end{split}
\end{equation}
Since $\lambda=0$ is the condition of local thermodynamic equilibrium, all partial derivatives at constant $\lambda$ are computed along the manifold of local equilibrium states. Hence, we can introduce the following notation:
\begin{equation}\label{hydro+tobulki}
\begin{split}
 \delta P_{\text{eq}} ={}& \dfrac{\partial P}{\partial \rho}\bigg|_{n,\lambda} \delta \rho + \dfrac{\partial P}{\partial n}\bigg|_{\rho,\lambda} \delta n  \, , \\
\delta \Pi ={}& \dfrac{\partial P}{\partial \lambda}\bigg|_{n,\rho} \delta \lambda \, , \\
\zeta ={}& \dfrac{\beta}{\Xi} \bigg( \dfrac{\partial P}{\partial \lambda}\bigg|_{\rho,n} \bigg)^2 \, , \\
\tau_\Pi ={}& \dfrac{1}{\Xi} \dfrac{\partial \phi}{\partial \lambda} \bigg|_{\rho,n} \, , \\
\end{split}
\end{equation}
and the last two equations of \eqref{comebemybaby} take the usual Israel-Stewart form, namely $(\rho+P)\partial_t \delta u_k +\partial_k  \delta (P_{\text{eq}}+\Pi) =0$ and $\tau_\Pi \partial_t \delta \Pi + \delta \Pi=-\zeta \partial_j \delta u^j$ (see Appendix \ref{AAA}). This completes the  ``Hydro+ $\rightarrow$ Israel-Stewart'' mapping. Note that no approximation has been made: the two theories are \textit{mathematically equivalent} in the linear regime. The physical connection between Hydro+ and a theory for bulk viscosity had already been noted (with a similar procedure as the one above) by \cite{Hydro+2018}. However, the full mathematical equivalence with Israel-Stewart theory had not been proved\footnote{Actually, \cite{Hydro+2018} argued that the Israel-Stewart theory is a particular case of Hydro+. Here, we have reversed the mapping: in the linear regime, it is always possible to rewrite Hydro+ as an Israel-Stewart theory. This is a manifestation of Theorem 3 of Ref.\ \cite{paperI}, according to which all thermodynamically consistent theories admit an ``Israel-Stewart representation''. Again, we would like to stress that this is a purely mathematical equivalence: Hydro+ and Israel-Stewart are governed by the same system of equations. The fact that the two theories were constructed having different domains of applicability in mind is irrelevant for our purposes.}. Indeed, the ``frequency-dependent bulk viscosity'' considered by \cite{Hydro+2018} coincides with the ``effective'' bulk viscosity coefficient of the Israel-Stewart theory that is observed in neutron-star models \cite{Sawyer_Bulk1989,BulkGavassino,
AlfordBulk}:
\begin{equation}
    \zeta_{\text{eff}} = \dfrac{\zeta}{1+\omega^2 \tau^2_\Pi} \, .
\end{equation}

\section{Some equivalent models in the linear regime}\label{sec:abacadabra}

In this section, we use Theorem 2 of our companion paper \cite{paperI} to show the mathematical equivalence (in the linear regime) of some fluid models. In a nutshell, we only need to show that there is a change of variables that maps the information current and entropy production rate of one model into the information current and entropy production rate of the other model. Then, it will automatically follow that the two models are mathematically equivalent, at least for linear deviations about homogeneous equilibrium.

\subsection{Fluid mixtures are equivalent to bulk viscosity} 

Consider a fluid mixture of two particle species, where one (say, baryons, $b$) has an associated conserved current, and the other (say, neutral pions, $\uppi^0$) does not have a corresponding conserved current. Mixtures of this kind are frequent in neutron stars \cite{Migdal1973,Suzuki1973,Haensel1982,
CamelioSimulations2022}. The first law of thermodynamics reads $d\rho= Tds+\mu db-\mathbb{A}d\uppi$, where $\rho$, $s$, $b$ and $\uppi$ are densities of energy, entropy, baryons, and pions, $T$ is the temperature, $\mu$ and $-\mathbb{A}$ are,  respectively, the baryon and the pion chemical potential. $\mathbb{A}$ is also the affinity \cite{PrigoginebookModernThermodynamics2014} of reactions like
\begin{equation}
b+b \ce{ <=> } b+b+\uppi^0 \, ,
\end{equation}
and it must vanish in thermodynamic equilibrium. Working in the equilibrium rest frame, the information current and entropy production rate for this system are \cite{GavassinoGibbs2021} 
\begin{equation}\label{infochemical}
\begin{split}
TE^0 ={}& \dfrac{1}{2} \bigg[\dfrac{\partial T}{\partial s}\bigg|_{b,\mathbb{A}} \! \! \! (\delta s)^2 + 2 \dfrac{\partial T}{\partial b}\bigg|_{s,\mathbb{A}} \! \! \! \delta s \, \delta b + \dfrac{\partial \mu}{\partial b}\bigg|_{s,\mathbb{A}} \! \! \!(\delta b)^2  -\dfrac{\partial \uppi}{\partial \mathbb{A}}\bigg|_{s,b} (\delta \mathbb{A})^2 + (\rho+P) \delta u^k \delta u_k  \bigg] \, , \\
TE^j ={}& \bigg[  \dfrac{\partial P}{\partial s}\bigg|_{b,\mathbb{A}} \delta s+ \dfrac{\partial P}{\partial b}\bigg|_{s,\mathbb{A}} \delta b+ \dfrac{\partial P}{\partial \mathbb{A}}\bigg|_{s,b} \delta \mathbb{A} \bigg] \delta u^j \, , \\
T\sigma ={}& \Xi \, (\delta \mathbb{A})^2 \, , \\
\end{split}
\end{equation}
where $P(s,b,\mathbb{A})$ is the total non-equilibrium pressure and $\Xi$ is the net pion production rate. All background quantities are evaluated at $\mathbb{A}=0$, which is the condition for chemical equilibrium. One can easily verify that, if we set $\varphi^A=\{\delta s, \delta b, \delta \mathbb{A}, \delta u^k \}$, and we write the system \eqref{Fudamental} explicitly, we obtain a system of field equations that is indeed equivalent to the linearized equations of motion of the mixture \cite{BulkGavassino}, but rewritten in a manifestly symmetric-hyperbolic form. Let us now introduce the following notation:
\begin{equation}
\begin{split}
\delta \Pi ={}& \dfrac{\partial P}{\partial \mathbb{A}}\bigg|_{s,b} \delta \mathbb{A} \, , \\
\zeta ={}& \dfrac{1}{\Xi} \bigg( \dfrac{\partial P}{\partial \mathbb{A}}\bigg|_{s,b} \bigg)^2 \, , \\
\tau_\Pi ={}& -\dfrac{1}{\Xi} \dfrac{\partial \uppi}{\partial \mathbb{A}} \bigg|_{s,b} \, . \\
\end{split}
\end{equation}
Then, Eq.\ \eqref{infochemical} becomes
\begin{equation}\label{infobulk}
\begin{split}
TE^0 ={}& \dfrac{1}{2} \bigg[\dfrac{\partial T}{\partial s}\bigg|_{b}  \! (\delta s)^2 + 2 \dfrac{\partial T}{\partial b}\bigg|_{s}  \! \delta s \, \delta b + \dfrac{\partial \mu}{\partial b}\bigg|_{s} \!(\delta b)^2 + \dfrac{\tau_\Pi}{\zeta} (\delta \Pi)^2 + (\rho+P) \delta u^k \delta u_k  \bigg] \, , \\
TE^j ={}& \bigg[  \dfrac{\partial P}{\partial s}\bigg|_{b} \delta s+ \dfrac{\partial P}{\partial b}\bigg|_{s} \delta b \bigg] \delta u^j + \delta \Pi \, \delta u^j \, , \\
T\sigma ={}& (\delta \Pi)^2/\zeta \, . \\
\end{split}
\end{equation}
These are the information current and entropy production rate of a bulk-viscous baryonic fluid, modeled using the Israel-Stewart theory \cite{Hishcock1983}, with bulk viscosity $\zeta$, relaxation time $\tau_\Pi$, and bulk-viscous stress $\delta \Pi$. In moving from equation \eqref{infochemical} to \eqref{infobulk}, we changed notation,
\begin{equation}
\dfrac{\partial f}{\partial g}\bigg|_{h,\mathbb{A}} = \dfrac{\partial f}{\partial g}\bigg|_{h} \, ,  
\end{equation} 
because the local-equilibrium equation of state $\rho(s,b)$ is obtained by imposing the constraint $\mathbb{A}=0$ in the non-equilibrium equation of state $\rho(s,b,\uppi)$, meaning that the Israel-Stewart thermodynamic derivatives are inevitably partial derivatives at constant $\mathbb{A}$ (with $\mathbb{A}=0$). 

We have explicitly constructed a mapping from the baryon-pion mixture to the Israel-Stewart theory for bulk viscosity, which satisfies the hypotheses of Theorem 2 of \cite{paperI}. Hence, in the linear regime, the baryon-pion mixture and Israel-Stewart bulk viscosity are mathematically equivalent hydrodynamic models. Indeed, this equivalence has already been observed, both theoretically \cite{BulkGavassino,GavassinoRadiazione,Camelio2022} and numerically \cite{CamelioSimulations2022}, but it had never been established with this level of mathematical rigor. In fact, not only these two systems ``contain the same physics''. In the linear regime, they are really the same system of equations. This implies that if we know the conditions for stability and causality of Israel-Stewart bulk viscosity, we also know those of the mixture. For example, the characteristic speed of Israel-Stewart bulk viscosity in the linear regime is \cite{Causality_bulk}
\begin{equation}
\begin{split}
c_{UV}^2 ={}& \dfrac{\partial P}{\partial \rho} \bigg|_{s/b} \!\! + \dfrac{\zeta}{\tau_\Pi (\rho+P)} = \\
& \dfrac{\partial P}{\partial \rho} \bigg|_{s/b,\mathbb{A}} \!\! \!\! -  \dfrac{\partial P}{\partial \mathbb{A}}\bigg|_{s,b} \dfrac{\partial P}{\partial \uppi}\bigg|_{s,b} \, \dfrac{1}{\rho+P} = \\
& \dfrac{\partial P}{\partial \rho} \bigg|_{s/b,\mathbb{A}} \!\! \!\! +  \dfrac{\partial P}{\partial \mathbb{A}}\bigg|_{\rho,s/b} \dfrac{\partial \mathbb{A}}{\partial \rho}\bigg|_{s/b,\uppi/b} \!\!= \dfrac{\partial P}{\partial \rho}\bigg|_{s/b,\uppi/b} \, , \\
\end{split}
\end{equation}
which is, indeed, the characteristic speed of the mixture \cite{Camelio2022}. Note that, to prove the equivalence between the two models, the explicit calculations performed here were not necessary. One could just note that the baryon-pion mixture belongs to the bulk viscosity universality class $(3,1,0)-(2,1,0)$ and, thus, a mapping of this kind \textit{must} exist. Also, note that, if we combine the results of this section with those of Section \ref{hydrIS}, we can conclude that Hydro+ can also be mapped into a two-component chemical mixture, and vice-versa (in the linear regime). This implies that, when we are close to equilibrium, there is no fundamental difference from an effective theory point of view between including a generic ``slow mode'' $\phi$ and adding an effective chemical reaction with a fictitious particle species \cite{BulkGavassino}.

\subsection{Superfluidity is equivalent to infinite heat conductivity}\label{entrainment} 

The relativistic version of Landau's two-fluid model  for superfluidity \cite{landau6,khalatnikov_book} is due to Carter \cite{carter1991,cool1995,langlois98} and Khalatnikov \cite{lebedev1982,Carter_starting_point,carter92}, see also \cite{Son2001,Gusakov2007,andersson2007review,
GavassinoIordanskii2021}. In the linear regime, the fields of the Carter-Khalatnikov theory are $\varphi^A=\{ \delta T, \delta \mu, \delta s^k, \delta n^k \}$, namely temperature, chemical potential, entropy flux, and particle flux. In the equilibrium rest frame, the information current and entropy production rate are \cite{GavassinoStabilityCarter2022,GavassinoCasmir2022} 
\begin{equation}\label{infosup}
\begin{split}
TE^0 ={}& \dfrac{1}{2} \bigg[\dfrac{\partial s}{\partial T}\bigg|_{\mu}  \! (\delta T)^2 + 2 \dfrac{\partial s}{\partial \mu}\bigg|_{T}  \! \delta T \, \delta \mu + \dfrac{\partial n}{\partial \mu}\bigg|_{T} \!(\delta \mu)^2 + \mathcal{K}^{ss}\delta s^k \delta s_k + 2\mathcal{K}^{sn}\delta s^k \delta n_k + \mathcal{K}^{nn}\delta n^k \delta n_k  \bigg] \, , \\
TE^j ={}& \delta T \delta s^j +\delta \mu \delta n^j \, , \\
T\sigma ={}& 0 \, , \\
\end{split}
\end{equation}
where $s$ and $n$ are the entropy and particle density. The phenomenological coefficients $\mathcal{K}^{XY}$, $X=s,n$, constitute the so-called ``entrainment matrix'', and they are related to the Landau ``superfluid mass density'' $\rho_S$ \cite{landau9} as follows:
\begin{equation}\label{entratevoi!}
\mathcal{K}^{XY}= 
\begin{bmatrix}
  \dfrac{1}{s^2}\bigg(Ts-\mu n + \dfrac{\mu^2n^2}{\rho_S}\bigg) & \, \, \dfrac{\mu}{s} \bigg( 1-\dfrac{\mu n}{\rho_S} \bigg)  \\
   \dfrac{\mu}{s} \bigg( 1-\dfrac{\mu n}{\rho_S} \bigg) &  \dfrac{\mu^2}{\rho_S} \\
\end{bmatrix} .
\end{equation}
Now, let us introduce the following change of notation:
\begin{equation}\label{beta1}
\begin{split}
\delta u^k ={}& \delta n^k/n \, ,  \\
\delta q^k ={}& T(\delta s^k-s\delta u^k) \, , \\
\beta_1 ={}&  \dfrac{1}{T^2 s^2}\bigg(Ts-\mu n + \dfrac{\mu^2n^2}{\rho_S}\bigg) \, , \\
\end{split}
\end{equation}
where $\delta u^k$ can be interpreted as the flow velocity (in the Eckart frame \cite{Eckart40}), and $\delta q^k$ can be interpreted as the heat flux.
Then, Eq.\ \eqref{infosup} becomes
\begin{equation}\label{infoheat}
\begin{split}
TE^0 ={}& \dfrac{1}{2} \bigg[\dfrac{\partial s}{\partial T}\bigg|_{\mu}  \! (\delta T)^2 + 2 \dfrac{\partial s}{\partial \mu}\bigg|_{T}  \! \delta T \, \delta \mu + \dfrac{\partial n}{\partial \mu}\bigg|_{T} \!(\delta \mu)^2 + (\rho+P)\delta u^k \delta u_k +2\delta u^k \delta q_k + \beta_1 \delta q^k \delta q_k  \bigg] \, , \\
TE^j ={}& (s  \delta T +n  \delta \mu)\delta u^j + \delta T \delta q^j/T \, , \\
T\sigma ={}& \delta q^k \delta q_k /\kappa T \spc  (\text{with }\kappa=+\infty) \, . \\
\end{split}
\end{equation}
We have recovered the information current and entropy production rate of the Israel-Stewart theory for heat conduction \cite{OlsonRegularCarter1990}, with infinite heat conductivity $\kappa$. Then, we can invoke Theorem 2 of \cite{paperI}, and we arrive at a surprisingly simple and intuitive result: in the linear regime, a relativistic superfluid is described by exactly the same equations of an ordinary heat-conducting fluid, in the limit of infinite heat conductivity\footnote{Note that the present equivalence only pertains to the dynamical equations of the system. To have a complete superfluid model, one also needs to impose a constraint on the initial conditions, by requiring that the superfluid momentum $\mathcal{K}^{nn}\delta n_j +\mathcal{K}^{ns}\delta s_j$ is irrotational \cite{cool1995}.}. This is due to the fact that superfluidity and heat conduction belong to ``adjacent'' universality classes, namely $(2,2,0)-(2,2,0)$ and $(2,2,0)-(2,1,0)$. Note, however, that the bridge is possible only within the Israel-Stewart approach to heat conduction. In fact, the ``second-order term'' $\beta_1 \delta q^k \delta q_k$ is crucial to the correspondence, as it carries all information about the value of $\rho_S$. At very low temperatures, if the thermal excitations are phononic, the third equation of \eqref{beta1} becomes  $\beta_1 = (Tsc_{\text{ph}}^2)^{-1}$, where $c_{\text{ph}}^2$ is the speed of phonons \cite{landau6}.

The mathematical equivalence outlined above also explains why the Israel-Stewart theory (and likewise the Cattaneo model \cite{cattaneo1958}) displays second sound \cite{rezzolla_book}: the structure of the equations is indistinguishable from that of Landau's two-fluid model for superfluidity which predicts, among other things, the existence of a second sound \cite{landau6}.

\subsection{Burgers viscoelasticity is equivalent to MIS$^*$} 

Consider a fluid at zero chemical potential whose perturbation-fields are $\varphi^A=\{\delta T,\delta u^k, \delta \Pi^{kl},\delta \Lambda^{kl}\}$, namely temperature, flow velocity, shear stresses, and an auxiliary two-tensor field. Both $\delta \Pi^{kl}$ and $\delta \Lambda^{kl}$ are symmetric and traceless. The most general information current and entropy production rate (up to field redefinitions \cite{GavassinoNonHydro2022}) are
\begin{equation}\label{infoBurgers}
\begin{split}
TE^0 ={}& \dfrac{c_v (\delta T)^2}{2T} + \dfrac{1}{2}(\rho{+}P)\delta u^k \delta u_k  +\dfrac{1}{2} \beta_2 (\delta \Pi^{kl}\delta \Pi_{kl}+\delta \Lambda^{kl}\delta \Lambda_{kl})\, , \\
TE^j ={}& s  \delta T \delta u^j + \delta \Pi^{jk}\delta u_k \, , \\
T\sigma ={}& \xi_1 \delta \Pi^{kl}\delta \Pi_{kl}+2\xi_2 \delta \Pi^{kl}\delta \Lambda_{kl}+\xi_3\delta \Lambda^{kl}\delta \Lambda_{kl} \, . \\
\end{split}
\end{equation}
If we write the field equations \eqref{Fudamental} explicitly, we recover the usual equations of a shear-viscous fluid in the Landau frame \cite{OlsonLifsh1990} with the only difference that, if $\xi_2 \neq 0$, the viscous stress $\delta \Pi^{kl}$ does not obey a relaxation equation as in the Israel-Stewart theory. Instead, it obeys a second-order equation,  
\begin{equation}\label{burgers}
(\lambda_2 \partial_t^2  + \lambda_1 \partial_t + 1) \delta \Pi_{kl} = - 2(\eta+\chi \partial_t) \braket{\partial_k \delta u_l} ,
\end{equation}
where $\braket{...}$ denotes the symmetric traceless part and
\begin{equation}
\begin{bmatrix}
\lambda_2 \\
\lambda_1 \\
2\eta \\
2\chi \\
\end{bmatrix}
= \dfrac{1}{\xi_1 \xi_3 - \xi_2^2} 
\begin{bmatrix}
\beta_2^2 \\
\beta_2 (\xi_1+\xi_3)\\
\xi_3 \\
 \beta_2 \\
\end{bmatrix} .
\end{equation}
In non-relativistic rheology, Eq.\  \eqref{burgers} is known as the ``Burgers model'' of viscoelasticity \cite{Findley_book,Prieto2014,Malek2018}. This describes a material that behaves like a fluid at low frequencies, with shear viscosity $\eta$, and like a solid at high frequencies, with shear modulus $G=\chi/\lambda_2=(2\beta_2)^{-1}$. The main difference with respect to Israel-Stewart theory is that the Burgers model presents two relaxation times, instead of just one. 

Assuming that $\xi_2 \neq 0$,\footnote{If $\xi_2=0$, then $\delta \Lambda_{kl}$ decouples from the other degrees of freedom, and the system becomes the Maxwell model (i.e. the Israel-Stewart theory). In this case, $\delta \Lambda_{kl}$ is an additional relaxing variable that has no influence on the mechanical properties of the material.} we can introduce the notation
\begin{equation}
\begin{split}
 z={}& \text{arcsinh}\bigg( \dfrac{\xi_3-\xi_1}{2\xi_2} \bigg),\\
\delta \Pi^{kl} ={}& \delta \Pi^{kl}_{(1)} + \delta \Pi^{kl}_{(2)} \, , \\
\delta \Lambda^{kl} ={}& e^z \, \delta \Pi^{kl}_{(1)} - e^{-z} \, \delta \Pi^{kl}_{(2)} \, , \\
2\eta_{(1)} ={}& (\xi_1+2\xi_2 e^z + \xi_3 e^{2z})^{-1}, \\
2\eta_{(2)} ={}& (\xi_1-2\xi_2 e^{-z} + \xi_3 e^{-2z})^{-1}, \\
\tau_{(1)}={}& 2\eta_{(1)}\beta_2 (1+e^{2z}) \, , \\
\tau_{(2)} ={}& 2\eta_{(2)}\beta_2 (1+e^{-2z})\, , \\
\end{split}
\end{equation} 
and \eqref{infoBurgers} becomes
\begin{equation}\label{infoKe}
\begin{split}
TE^0 ={}& \dfrac{c_v (\delta T)^2}{2T} + \dfrac{1}{2}(\rho{+}P)\delta u^k \delta u_k +\dfrac{\tau_{(1)}}{4 \eta_{(1)}} \delta \Pi^{kl}_{(1)}\delta \Pi_{(1)kl}+\dfrac{\tau_{(2)}}{4 \eta_{(2)}} \delta \Pi^{kl}_{(2)}\delta \Pi_{(2)kl}\, , \\
TE^j ={}& s  \delta T \delta u^j + \big(\delta \Pi^{jk}_{(1)}+\delta \Pi^{jk}_{(2)} \big)\delta u_k \, , \\
T\sigma ={}&   \dfrac{\delta\Pi^{kl}_{(1)}\delta \Pi_{(1)kl}}{2\eta_{(1)}}+\dfrac{\delta \Pi^{kl}_{(2)}\delta \Pi_{(2)kl}}{2\eta_{(2)}} \, . \\
\end{split}
\end{equation}
These are the information current and entropy production rate of the MIS$^*$ theory, formulated by \cite{KeYin2022} as a model of the quark-gluon plasma in the extended hydrodynamic regime. Indeed, if we write \eqref{Fudamental} explicitly, we recover the linearized field equations of MIS$^*$, expressed in a symmetric-hyperbolic form. In a nutshell, MIS$^*$ models a shear-viscous fluid, whose shear stress tensor is the sum of two parts, $\delta \Pi_{(1)}^{kl}$ and $\delta \Pi^{kl}_{(2)}$, subject to two independent relaxation equations:
\begin{equation}\label{KeYinequations}
\begin{split}
(\tau_{(1)}\partial_t + 1)\delta \Pi_{(1)kl} ={}& -2\eta_{(1)} \braket{\partial_k \delta u_l} \, , \\
(\tau_{(2)}\partial_t + 1)\delta \Pi_{(2)kl} ={}& -2\eta_{(2)} \braket{\partial_k \delta u_l} \, .\\
\end{split}
\end{equation}
By Theorem 2 of our companion paper \cite{paperI}, we can conclude that the Burgers model and MIS$^*$ are mathematically equivalent (in the linear regime), and they represent equally well the universality class $(1,1,2)-(1,1,0)$. This equivalence implies, for example, that MIS$^*$ features an elastic behavior at high frequencies. More importantly, it implies that, if $\xi_2 \neq 0$, the Burgers equation \eqref{burgers} can \textit{always} be decomposed in the form \eqref{KeYinequations}, independently from whether we are able to assign a microscopic meaning to the individual contributions $\delta \Pi_{(1)}^{kl}$ and $\delta \Pi^{kl}_{(2)}$. In fact,  \eqref{KeYinequations} can be viewed as a convenient reparameterization of the Burgers degrees of freedom. The four parameters in \eqref{burgers} can be expressed in terms of the parameters in \eqref{KeYinequations} as follows:
\begin{equation}\label{ultimatum}
\begin{split}
\lambda_2 ={}& \tau_{(1)}\tau_{(2)}  \, ,\\
\lambda_1 ={}& \tau_{(1)}+\tau_{(2)} \, ,  \\
\eta ={}& \eta_{(1)}+\eta_{(2)} \, ,  \\
\chi ={}& \eta_{(1)} \tau_{(2)}+\eta_{(2)}\tau_{(1)} \, . \\
\end{split}
\end{equation}
If we plug \eqref{ultimatum} into \eqref{burgers}, and we introduce the partial shear moduli $G_{(1)}=\eta_{(1)}/\tau_{(1)}$ and $G_{(2)}=\eta_{(2)}/\tau_{(2)}$ \cite{BaggioliHolography}, we obtain
\begin{equation}
\dfrac{\eta_{(1)}\eta_{(2)}}{G_{(1)}G_{(2)}} \partial_t^2 \delta \Pi_{kl} + \bigg(\dfrac{\eta_{(1)}}{G_{(1)}} +\dfrac{\eta_{(2)}}{G_{(2)}} \bigg) \partial_t \delta \Pi_{kl} + \delta \Pi_{kl}  {=} {-} 2(\eta_{(1)}{+}\eta_{(2)}) \! \braket{\partial_k \delta u_l} {-}2 \dfrac{\eta_{(1)}\eta_{(2)} (G_{(1)}\!{+}G_{(2)})}{G_{(1)}G_{(2)}} \partial_t \! \braket{\partial_k \delta u_l} \! ,
\end{equation}
which in non-relativistic rheology is a well-known  ``representation'' of the Burgers model \cite{Malek2018}.

\section{A priori construction of the Israel-Stewart class}\label{sec:apriori}

Until this point we have considered examples of existing theories in the literature and we were using Theorems 1 and 2 of our companion paper \cite{paperI} as a means to study their mathematical structure. However, it is perhaps more interesting to work the other way around, and use Theorems 1, 2, and 3  to construct the most general linear theory for a given universality class. Here we provide an explicit example. We will show that the linearized Israel-Stewart theory in the Eckart frame \cite{Israel_Stewart_1979} is the most general theory of universality class $(3,2,1)-(2,1,0)$ having a symmetric stress-energy tensor.

\subsection{Most general information current and entropy production rate}

Theorems 1 and 2 of \cite{paperI} tell us that, once we have fixed the triplet $\{ \varphi^A,E^\mu,\sigma\}$, the equations of the theory are fully determined. Hence, the first step is to construct the most general $E^\mu$ and $\sigma$, using 6 fields, $\varphi^A =\{\delta \mu,\delta T ,\delta u^k,\delta \Pi,\delta q^k, \delta \Pi^{jk} \}$, where 3 of them ($\delta \mu$, $\delta T$, and $\delta \Pi$) are  scalars under spatial rotations, 2 are vectors ($\delta u^k$, and $\delta q^k$) and the remaining one ($\delta \Pi^{jk}$) is a symmetric traceless  tensor with two indices. For the time being, the fields $\varphi^A$ do not carry any precise physical meaning. This means that, e.g., ``$\, \delta T \,$'' is just the name given to a field, which may or may not coincide with the temperature displacement. The most general quadratic functions $E^0$, $E^j$, and $\sigma$ that are compatible with rotational invariance (recall that the equilibrium state is assumed isotropic) are given below:
\begin{equation}\label{zarrillo}
\begin{split}
TE^0={}& \dfrac{1}{2} \bigg[A_1 (\delta \mu)^2+2A_2 \delta \mu \delta T+A_3 (\delta T)^2+2A_4 \delta T \delta \Pi +A_5 (\delta \Pi)^2 +2A_6 \delta \Pi \delta \mu \\
 & + B_1 \delta u^k \delta u_k + 2 B_2 \delta u^k \delta q_k + B_3 \delta q^k \delta q_k + C_1 \delta \Pi^{jk} \delta \Pi_{jk} \bigg] \, ,  \\
TE^j={}& (K_1 \delta \mu +K_2 \delta T +K_3 \delta \Pi)\delta u^j +(H_1 \delta \mu +H_2 \delta T +H_3 \delta \Pi)\delta q^j + (I_1 \delta u_k +I_2 \delta q_k)\delta \Pi^{jk} \, ,  \\
T\sigma ={}& V_1 (\delta \mu)^2+2V_2 \delta \mu \delta T+V_3 (\delta T)^2+2V_4 \delta T \delta \Pi +V_5 (\delta \Pi)^2 +2V_6 \delta \Pi \delta \mu \\
 & + W_1 \delta u^k \delta u_k + 2 W_2 \delta u^k \delta q_k + W_3 \delta q^k \delta q_k + Z_1 \delta \Pi^{jk} \delta \Pi_{jk} \, . \\
\end{split}
\end{equation}
Equation \eqref{zarrillo} presents in total $28$ free parameters. However, different choices of such parameters do not necessarily give rise to different theories. In fact, according to Theorem 2 \cite{paperI}, two theories are equivalent if and only if there is a field redefinition that maps the information current and the entropy production rate of the first theory into the information current and the entropy production rate of the second. In our case, the most general field redefinitions that preserve rotational invariance are reported below:
\begin{equation}\label{redefine}
\begin{split}
\delta \tilde{\mu} ={}& h_1 \delta \mu +h_2 \delta T +h_3 \delta \Pi \, , \\
\delta \tilde{T} ={}& h_4 \delta \mu +h_5 \delta T +h_6 \delta \Pi \, ,  \\
 \delta \tilde{\Pi} ={}& h_7 \delta \mu +h_8 \delta T +h_9 \delta \Pi \, ,  \\
 \delta \tilde{u}^k ={}& l_1 \delta u^k + l_2 \delta q^k \, ,  \\
 \delta \tilde{q}^k ={}& l_3 \delta u^k + l_4 \delta q^k \, ,  \\
 \delta \tilde{\Pi}^{jk} ={}& m_1 \delta \Pi^{jk} \, . \\
\end{split}
\end{equation}
Let us use this freedom of redefining fields to our advantage. We follow the same steps as in the proof of Theorem 3 of \cite{paperI}. 

For the sake of clarity, let us interpret the conservation laws respectively as  particles, energy, and momentum. Then, the linear theory presents some stationary states, which correspond to neighboring homogeneous thermodynamic equilibrium states. Such states can be parameterized using 5 numbers: 
\begin{equation}
  \mu^I=\{\text{``displacement in chemical potential''},\text{``displacement in temperature''}, \text{``displacement in flow-velocity''}\}\, .  
\end{equation}
The value of the fields $\varphi^A$ in such states can be expressed (by linearity of the theory) as a linear combination of the thermodynamic displacements: $\varphi^A = \mathcal{N}^A_I \mu^I$. Let us interpret this formula within a linear algebra perspective. At a spacetime event $p$, all possible values of $\varphi^A(p)$ form a vector space, with dimension $14$. Within such vector space, we can identify a 5-dimensional vector subspace of local equilibrium states, and the five vectors $\mathcal{N}_I^A$ form a basis of such subspace. Then, we can extend $\mathcal{N}^A_I$ to a basis of the full vector space, by introducing other 9 vectors $\mathcal{N}_a^A$ (which can be easily chosen so that rotation invariance is preserved). This implies that an arbitrary state can be expressed in the form $\varphi^A=\mathcal{N}^A_I \mu^I +\mathcal{N}^A_a  \Pi^a$, where $\{\mu^I, \Pi^a \}$ are 14 linear combination coefficients. Since the expansion of a vector on a basis is always unique, we can invert the relation above, and write $\mu^I = \mathcal{G}^I_A \varphi^A$ and $\Pi^a = \mathcal{G}^a_A \varphi^A$. But this can be viewed as a field redefinition, $\tilde{\varphi}^C = \mathcal{G}^C_A \varphi^A$, with $\tilde{\varphi}^C= \{\mu^I ,\Pi^a \}$. In conclusion, it is possible to perform a field redefinition such that the new fields  $\tilde{\varphi}^C =\{\delta \mu,\delta T ,\delta u^k,\delta \Pi,\delta q^k, \delta \Pi^{jk} \}$ reduce to $\{\hat{\mu}-\mu,\hat{T}- T,\hat{u}^k-u^k,0,0, 0 \}$ whenever the perturbation connects our reference equilibrium state $\{\mu,T,u^k\}$ to a neighboring equilibrium state $\{ \hat{\mu},\hat{T},\hat{u}^k \}$. But for perturbations of this kind, the fluid is indistinguishable from a perfect fluid, for which we have \cite{GavassinoCausality2021}
\begin{equation}\label{tretisette}
\begin{split}
 TE^0 ={}& \dfrac{1}{2} \bigg[ \dfrac{\partial n}{\partial \mu}\bigg|_T (\delta \mu)^2+ 2\dfrac{\partial n}{\partial T}\bigg|_\mu \delta \mu \, \delta T + \dfrac{\partial s}{\partial T}\bigg|_\mu (\delta T)^2 + (\rho+P)\delta u^k \delta u_k \bigg] \, , \\
TE^j ={}& (n\, \delta \mu  + s \, \delta T) \delta u^j \, ,\\
T\sigma ={}& 0 \, .\\
\end{split}
\end{equation}
It follows that, for the particular choice of fields $\tilde{\varphi}^B$ introduced above, equation \eqref{zarrillo} simplifies considerably:
\begin{equation}
\begin{split}
& A_1 (\delta \mu)^2+2A_2 \delta \mu \delta T+A_3 (\delta T)^2=\delta n \delta \mu + \delta s \delta T \, , \\
& B_1-(\rho+P)=K_1-n=K_2-s = V_1=V_2=V_3=W_1=0 \, . \\
\end{split}
\end{equation}
However, since $\sigma$ must be non-negative definite, if $V_1$, $V_3$, and $W_1$ vanish, we need also to set $V_4=V_6=W_2=0$.  Therefore, \eqref{zarrillo} becomes
\begin{equation}\label{zarrillo2}
\begin{split}
TE^0={}& \dfrac{1}{2} \bigg[\delta n \delta \mu + \delta s \delta T+2A_4 \delta T \delta \Pi +A_5 (\delta \Pi)^2 +2A_6 \delta \Pi \delta \mu  + (\rho{+}P) \delta u^k \delta u_k + 2 B_2 \delta u^k \delta q_k + B_3 \delta q^k \delta q_k + C_1 \delta \Pi^{jk} \delta \Pi_{jk} \bigg] \, ,  \\
TE^j={}& (\delta P+ K_3 \delta \Pi)\delta u^j +(H_1 \delta \mu +H_2 \delta T +H_3 \delta \Pi)\delta q^j + ( I_1 \delta u_k +I_2 \delta q_k)\delta \Pi^{jk} \, ,  \\
T\sigma ={}&  V_5 (\delta \Pi)^2 +
   W_3 \delta q^k \delta q_k + Z_1 \delta \Pi^{jk} \delta \Pi_{jk} \, . 
\end{split}
\end{equation}
But the information current can be simplified even further. In fact, the requirement that $\mu +\delta \mu$, $T+\delta T$, and $\delta u^k$ are respectively the chemical potential, the temperature, and the flow-velocity of neighboring equilibrium states does not fix the non-equilibrium fields $\delta \mu$, $\delta T$ and $\delta u^k$ completely. There are still some field redefinitions that preserve this condition, namely $\delta \tilde{\mu} = \delta \mu +h_3 \delta \Pi$, $\delta \tilde{T} = \delta T +h_6 \delta \Pi$, and $\delta \tilde{u}^k = \delta u^k + l_2 \delta q^k$. Such field redefinitions constitute a ``change of hydrodynamic frame'', and can be used to our advantage. In fact, with a change of hydrodynamic frame, we can always redefine the flow-velocity so that $H_1=0$. We can also redefine temperature and chemical potential so that $A_4=A_6=0$. This fixes the hydrodynamic frame completely (Eckart frame). Finally, we can also rescale $\delta q^k$, $\delta \Pi$, and $\delta \Pi^{jk}$ (with some transformations $\delta \tilde{q}^k=l_4 \delta q^k$, $\delta \tilde{\Pi} = h_9 \delta \Pi$, and $\delta \tilde{\Pi}^{jk} = m_1 \delta \Pi^{jk} $) in such a way that $B_2=K_3=I_1=1$,\footnote{By doing this, we are implicitly assuming that  $B_2$, $K_3$, and $I_1$ are not zero. The special cases where one or more of these coefficients vanish may be seen as ``singular limits'' of the theory that we are constructing here.} and Eq.\ \eqref{zarrillo2} reduces to
\begin{equation}\label{zarrillo3}
\begin{split}
TE^0={}& \dfrac{1}{2} \bigg[\delta n \delta \mu + \delta s \delta T +A_5 (\delta \Pi)^2  + (\rho{+}P) \delta u^k \delta u_k + 2 \delta u^k \delta q_k + B_3 \delta q^k \delta q_k + C_1 \delta \Pi^{jk} \delta \Pi_{jk} \bigg] \, ,  \\
TE^j={}& (\delta P+ \delta \Pi)\delta u^j +(H_2 \delta T +H_3 \delta \Pi)\delta q^j + ( \delta u_k +I_2 \delta q_k)\delta \Pi^{jk} \, ,  \\
T\sigma ={}&  V_5 (\delta \Pi)^2 +
   W_3 \delta q^k \delta q_k + Z_1 \delta \Pi^{jk} \delta \Pi_{jk}  \, . \\
\end{split}
\end{equation}
Note what we have just shown: with an appropriate field redefinition, it is always possible to recast \eqref{zarrillo} in the form \eqref{zarrillo3}. The expressions for $E^\mu$ and $\sigma$ given above coincide with the information current and the entropy production rate of the Israel-Stewart theory in the Eckart frame \cite{Hishcock1983}, with only one difference: the factor $H_2$, which in the Israel-Stewart theory equals $T^{-1}$, here is completely free. However, if we write the field equations $E^\mu_{AB}\partial_\mu \varphi^B=-\sigma_{AB}\varphi^B$ explicitly, we can compute the conservation laws of the system (see, e.g., \cite{GavassinoNonHydro2022}). This allows us to derive the (linearized) formulas for all the components of the stress-energy tensor. For example, with the above choice of $E^\mu$ and $\sigma$, the displacement in energy current and the displacement in momentum density \cite{MTW_book} are, respectively,
\begin{equation}
    \begin{split}
        \delta T^{j0} ={}& (\rho+P)\delta u^j +TH_2\delta q^j \, , \\
        \delta T^{0j} ={}& (\rho+P)\delta u^j +\delta q^j \, . \\
    \end{split}
\end{equation}
Hence, if we additionally require that the stress-energy tensor be symmetric, we immediately obtain $H_2=T^{-1}$, and the Israel-Stewart information current is fully recovered.


\subsection{Implications}

The above analysis has far-reaching consequences. It implies that, if the degrees of freedom of a hydrodynamic theory have the same geometric character (here, the same transformation laws under spatial rotations) as those of the Israel-Stewart theory, and the conservation laws are those of the Israel-Stewart theory, then in the linear regime such theory becomes indistinguishable from the Israel-Stewart theory in the Eckart frame, if the hypotheses of Theorem 1 \cite{paperI} are met. This implies, for example, that within the Israel-Stewart framework all hydrodynamic frames must be mathematically equivalent in the linear regime \cite{GavassinoNonHydro2022} (this does not include the Israel-Stewart theory in the ``general frame'' \cite{NoronhaGeneralFrame2021}, because the latter has more degrees of freedom). It also implies that divergence-type theories \cite{Geroch1995}, Carter's theory for viscosity \cite{carter1991}, BRSSS \cite{Baier2008}, DNMR \cite{Denicol2012Boltzmann} and the GENERIC theory \cite{Stricker2019} must all become indistinguishable from the Israel-Stewart theory, near equilibrium. Indeed, all the second-order theories for viscosity belong to the same universality class: $(3,2,1)-(2,1,0)$.

\section{Relativistic supersolid hydrodynamics}\label{sec:supersolid}

To illustrate the predictive power of the formalism, let us use it to construct (with essentially no effort) a linearized model for supersolid hydrodynamics. Note that what follows is the first linear model of a relativistic supersolid which is proved to be causal, stable, and strongly hyperbolic (for other approaches, see \cite{Pelemetrski2009,NicolisSupersolid2014}).

In the linear regime, an isotropic supersolid can be interpreted as a system of class $(3,2,1)-(3,2,1)$, whose fields are $\mu^I=\{\delta T, \delta \mu, \delta \Pi, \delta s^k, \delta n^k, \delta \Pi^{kl} \}$, which may be interpreted as the perturbations to respectively temperature, chemical potential, elastic bulk stress, entropy flux, particle flux, and elastic shear stress. No dissipative effects are considered here (i.e. there are as many fields as conservation laws). Thus, the supersolid information current can be easily guessed by combining the superfluid information current \eqref{infosup} with the elastic information current (see companion paper \cite{paperI}). The result is the following:
\begin{equation}\label{infosupsol}
\begin{split}
TE^0 ={}& \dfrac{1}{2} \bigg[\dfrac{\partial s}{\partial T}\bigg|_{\mu}  \! \! \! (\delta T)^2 {+} 2 \dfrac{\partial s}{\partial \mu}\bigg|_{T}  \! \! \! \delta T \, \delta \mu {+} \dfrac{\partial n}{\partial \mu}\bigg|_{T}  \! \! \!(\delta \mu)^2 + \mathcal{K}^{ss}\delta s^k \delta s_k + 2\mathcal{K}^{sn}\delta s^k \delta n_k + \mathcal{K}^{nn}\delta n^k \delta n_k + \dfrac{(\delta \Pi)^2}{K} {+} \dfrac{\delta \Pi^{kl}\delta \Pi_{kl}}{2G} \bigg]  , \\
TE^j ={}& \delta T \delta s^j +\delta \mu \delta n^j + s^{-1} \delta \Pi \delta s^j + s^{-1} \delta \Pi^{jk} \delta s_k \, , \\
T\sigma ={}& 0 \, , \\
\end{split}
\end{equation}
where $s$ is the entropy density, $n$ is the particle density, $\mathcal{K}^{XY}$ is the entrainment matrix (see section \ref{entrainment}), $K$ is the bulk modulus, and $G$ is the shear modulus. In the formula for $TE^j$, we postulated the coupling $ s^{-1} \delta \Pi \delta s^j + s^{-1} \delta \Pi^{jk} \delta s_k$, because in the Andreev-Lifshitz theory \cite{Andreev1969} only the normal component participates in the elastic behavior, and $\delta u_N^j:=\delta s^j/s$ is (by definition) the velocity of the normal flow \cite{Termo,GavassinoKhalatnikov2022}. Note that, in the fully non-linear supersolid theory, the elastic stress $\Pi^{kl} +\Pi \, \delta^{kl}$ is not symmetric. However, it becomes symmetric in the linear regime \cite{Andreev1969}, so that (as usual) $\delta \Pi^{kl}$ is assumed symmetric and traceless. 
The field equations \eqref{Fudamental}, computed from \eqref{infosupsol}, read 
\begin{equation}\label{supersoliffieldeq}
\begin{split}
\partial_t \bigg[\dfrac{\partial s}{\partial T}\bigg|_{\mu}   \delta T + \dfrac{\partial s}{\partial \mu}\bigg|_{T}   \delta \mu  \bigg] +\partial_j \delta s^j ={}& 0 \, , \\
\partial_t \bigg[\dfrac{\partial n}{\partial \mu}\bigg|_{T}  \delta \mu + \dfrac{\partial s}{\partial \mu}\bigg|_{T}   \delta T  \bigg] +\partial_j \delta n^j={}& 0 \, , \\
\partial_t (\mathcal{K}^{ss} \delta s_k + \mathcal{K}^{sn}\delta n_k) +\partial_k (\delta T+s^{-1} \delta \Pi) + s^{-1} \partial_j \delta \Pi^j_k={}& 0 \, , \\
\partial_t (\mathcal{K}^{nn} \delta n_k + \mathcal{K}^{sn}\delta s_k) +\partial_k \delta \mu={}& 0 \, , \\
K^{-1} \partial_t \delta \Pi +\partial_j (s^{-1} \delta s^j)={}& 0 \, , \\
(2G)^{-1} \partial_t \delta \Pi_{kl} + \braket{\partial_k (s^{-1}\delta s_l)}={}& 0 \, , \\
\end{split}
\end{equation}
and they constitute a manifestly symmetric hyperbolic system. Our goal, now, is to show that this system is indeed the correct relativistic generalization of supersolid hydrodynamics \cite{Sears2010} (in the non-dissipative limit). 

First, we see that, since by definition $\delta s^j=s\delta u_N^j$, the first equation of \eqref{supersoliffieldeq} simplifies to $\partial_t \delta s+\partial_j (s\delta u_N^j)=0$, i.e. the entropy is advected by the normal component, compare with Eqs.\ (8) and (13) of \cite{Sears2010} (neglecting dissipation). Also, if we introduce the lattice displacement $\xi^k$, whose defining equation is $\partial_t \xi^k = \delta u_N^k$ (see Eqs.\ (9) and (14) of \cite{Sears2010}), the last two equations of \eqref{supersoliffieldeq} can be integrated to give $\delta \Pi=-K\partial_j \xi^j$, and $\delta \Pi_{kl}=-2G\braket{\partial_k \xi_l}$, see eq.s (22) and (43) of \cite{Sears2010}. Furthermore, we know from Carter's multifluid theory \cite{cool1995} that the Landau superfluid velocity $\delta u_S^j$ is defined through the equation $\mu  \delta u_S^j = \mathcal{K}^{nn}\delta n^j+\mathcal{K}^{sn}\delta s^j$ (note that $\mu$ is the \textit{relativistic} chemical potential: It contains the ``$mc^2$'' term). As a consequence, the fourth equation of \eqref{supersoliffieldeq} becomes $\mu \partial_t \delta u_{Sk}+\partial_k \delta \mu=0$, which is the relativistic analogue of eq.s (12) and (16) of \cite{Sears2010}. One should keep in mind that, in the Newtonian limit, $\mu \rightarrow m$, where $m$ is the mass of the constituents. Additionally, note that what \cite{Sears2010} call ``$\mu$'' is actually $\mu/m -1$ for us.

Thanks to the well-known Maxwell relation
\begin{equation}
    \dfrac{\partial s}{\partial \mu}\bigg|_{T}  = \dfrac{\partial n}{\partial T}\bigg|_{\mu} \, ,
\end{equation}
the second equation of \eqref{supersoliffieldeq} reduces to the usual continuity equation, $\partial_t \delta n+\partial_j \delta n^j=0$, see eq. (10) of \cite{Sears2010}. However, one needs to be careful here, because in relativity the momentum density, $\delta T^{0j}$, and the current of rest mass, $m \delta n^j$, do not coincide. Instead,  $\delta T^{0j}$ is a linear combination of both $\delta n^j$ and $\delta s^j \,$:
\begin{equation}\label{totalmomentum}
    \delta T^{0j} = (n\mathcal{K}^{nn}+s\mathcal{K}^{sn})\delta n^j +(n\mathcal{K}^{sn}+s\mathcal{K}^{ss})\delta s^j = \mu \delta n^j +T \delta s^j = \rho_S \delta u_S^j+\rho_N \delta u_N^j \, ,
\end{equation}
where $\rho_S$ is Landau's superfluid mass density, defined through equation \eqref{entratevoi!}, and $\rho_N$ is Landau's normal mass density, defined through the condition $sT+\mu n=\rho_S+\rho_N$ (relativistic partition of inertia \cite{GavassinoStabilityCarter2022}). The first equality sign in \eqref{totalmomentum} follows from multifluid hydrodynamics \cite{cool1995}, the second follows from \eqref{entratevoi!}, and the third follows from the definitions of $\delta u_S^j$ and $\delta u_N^j$. Combining the third and fourth equations of \eqref{supersoliffieldeq}, using the identity $\delta P =s\delta T +n \delta \mu$, we obtain $\partial_t \delta T^0_k+\partial_k (\delta P+\delta \Pi)+\partial_j\delta \Pi^j_k=0$, which matches Eqs.\ (4), (11), and (15) of \cite{Sears2010}, thus completing our comparison.

Causality, stability, and strong hyperbolicity for the present model are guaranteed, if the vector field $E^\mu$ is future directed timelike for all non-vanishing perturbations. On the other hand, we can use the mapping developed in Section \ref{entrainment} to transform the information current \eqref{infosupsol} into that of the Israel-Stewart theory in the Eckart frame, as given in   \eqref{zarrillo3}. It follows that the conditions for causality, stability, and strong hyperbolicity are exactly those of the Israel-Stewart theory, see Eqs.\ (53)-(60) of \cite{Hishcock1983}.

\section{Conclusions}
\label{conclusions}

This article, together with its companion \cite{paperI}, paves the way to a new era of relativistic hydrodynamics, where different theories are distinguished, not by acronyms, but by the mechanical properties of the materials they describe (at least in the linear regime). With a rather diverse list of examples, we hope we have convinced the reader that, despite the hydrodynamic landscape is immense  \cite{Israel_Stewart_1979,Baier2008,Denicol2012Boltzmann,Florkowski:2010cf,Martinez:2010sc,Jaiswal:2013vta,Bemfica2017TheFirst,Kovtun2019,Bemfica:2019knx,Hoult:2020eho,Bemfica-Disconzi-Graber-2021,BemficaDNDefinitivo2020,Florkowski_2018,
RauWasserman2020,Kiamari_2021,AmmonChiral2021,
SperanzaChiral2021,Hongo2021,
NoronhaGeneralFrame2021,Perna_2021,
GavassinoKhalatnikov2022,KeYin2022,
Singh2022,Most_2022,
HellerSingulant2022,Brito_2022,Wagner:2022ayd,SalazarZannias2022,Bhadury2022}, there is still hope (at least in the linear regime) for a systematic and rigorous classification of theories into universal ``rheological models''. The reason why this is possible is that the information current $E^\mu$ and the entropy production rate $\sigma$ play for linearized hydrodynamics a role that is comparable to that of the Lagrangian density in classical and quantum field theory: they contain full information about the equations of the theory. Hence, just like in field theory, one builds the most general Lagrangian density compatible with the symmetries of the system, similarly in hydrodynamics one can build the most general information current and entropy production rate compatible with the conservation laws, and this automatically produces the most general (thermodynamically consistent) theory for a given set of degrees of freedom. 

We believe that the examples that we chose here are only the ``tip of the iceberg'' of what may be done with the present formalism. In fact, to keep the discussion simple, we did not consider theories whose information flux $TE^j$ contains couplings of the form $\varepsilon\indices{^j _k _l}\varphi^k \psi^l$. The inclusion of such terms would allow us to enter the world of magnetohydrodynamics \cite{anile_1990,Denicol:2018rbw,Most:2021uck,Most:2021rhr}, spin hydrodynamics \cite{Florkowski_2018,Hongo2021,Bhadury2022}, and chiral hydrodynamics \cite{AmmonChiral2021,SperanzaChiral2021}, where interesting mappings and equivalences are still there to be discovered. Finally, it would be fascinating to determine what remains from these universality classes as nonlinear corrections are taken into account. We leave such questions to future investigations.

\section*{Acknowledgements}
M.M.D. is partially supported by NSF grant DMS-2107701, a Vanderbilt's Seeding Success Grant, 
a Chancellor's Faculty Fellowship, and DOE grant DE-SC0024711.
L.G. is partially supported by a Vanderbilt's Seeding Success Grant. J.N. is partially supported by the U.S. Department of Energy, Office of Science, Office for Nuclear Physics
under Award No. DE-SC0023861.

\appendix

\section{Bulk viscous behavior of Hydro+}\label{AAA}

The fourth equation of \eqref{comebemybaby}, expressed in terms of $\delta \rho$, $\delta n$, and $\delta \lambda$, reads
\begin{equation}\label{quebuquenz}
\dfrac{\partial \phi}{\partial \rho}\bigg|_{n,\lambda} \partial_t \delta \rho + \dfrac{\partial \phi}{\partial n}\bigg|_{\rho,\lambda} \partial_t\delta n +\dfrac{\partial \phi}{\partial \lambda}\bigg|_{n,\rho} \partial_t\delta \lambda + A_\phi \partial_j \delta u^j = -\Xi \delta \lambda \, .
\end{equation}
Using the first two equations of \eqref{comebemybaby}, we can rearrange \eqref{quebuquenz} as follows:
\begin{equation}\label{nonancorabellabella}
\dfrac{1}{\Xi}\dfrac{\partial \phi}{\partial \lambda}\bigg|_{n,\rho} \partial_t\delta \lambda + \delta \lambda =\dfrac{1}{\Xi} \bigg[(\rho+P)\dfrac{\partial \phi}{\partial \rho}\bigg|_{n,\lambda}  + n \dfrac{\partial \phi}{\partial n}\bigg|_{\rho,\lambda}-A_\phi \bigg]\partial_j \delta u^j \, .
\end{equation}
Now, using \eqref{deppummo}, we can derive a simple thermodynamic relation \cite{Hydro+2018} (which holds only at equilibrium):
\begin{equation}\label{gabbuzzzo}
\beta \dfrac{\partial P}{\partial \lambda}\bigg|_{\rho,n} = -(\rho+P) \dfrac{\partial \beta}{\partial \lambda}\bigg|_{\rho,n} + n \dfrac{\partial \alpha}{\partial \lambda}\bigg|_{\rho,n} +A_\phi \, .
\end{equation}
Furthermore, defined the Massieu function $\mathcal{L}=s+\lambda \phi$, we have the differential $d\mathcal{L}=\beta \, d\rho -\alpha \, dn + \phi \, d \lambda$, from which we can derive two useful Maxwell relations:
\begin{equation}
\dfrac{\partial \beta}{\partial \lambda}\bigg|_{\rho,n}=\dfrac{\partial \phi}{\partial \rho}\bigg|_{n,\lambda} \, ,  \spc \dfrac{\partial \alpha}{\partial \lambda}\bigg|_{\rho,n}=-\dfrac{\partial \phi}{\partial n}\bigg|_{n,\lambda} \, . 
\end{equation}
This allows us to rewrite equation \eqref{gabbuzzzo} as 
\begin{equation}
-\beta \dfrac{\partial P}{\partial \lambda}\bigg|_{\rho,n} = (\rho+P)\dfrac{\partial \phi}{\partial \rho}\bigg|_{n,\lambda} + n \dfrac{\partial \phi}{\partial n}\bigg|_{n,\lambda} -A_\phi \, ,
\end{equation}
and \eqref{nonancorabellabella} becomes
\begin{equation}
\dfrac{1}{\Xi}\dfrac{\partial \phi}{\partial \lambda}\bigg|_{n,\rho} \partial_t\delta \lambda + \delta \lambda =-\dfrac{\beta}{\Xi} \dfrac{\partial P}{\partial \lambda}\bigg|_{\rho,n} \partial_j \delta u^j \, .
\end{equation}
Multiplying both sides by $\partial P/\partial \lambda$ (at constant $\rho$, and $n$) we finally obtain $\tau_\Pi \partial_t \delta \Pi+\delta \Pi=-\zeta \partial_j \delta u^j$, where $\delta \Pi$, $\tau_\Pi$ and $\zeta$ are defined in equation \eqref{hydro+tobulki}.

\bibliography{Biblio}

\label{lastpage}

\end{document}